\documentclass[doublecol]{epl2}
\usepackage{graphicx}

\title{Phase Transitions and Chaos in Long-Range Models of Coupled Oscillators}
\shorttitle{Title} 

\author{G.Miritello\inst{1} \and A. Pluchino\inst{1} \and A. Rapisarda\inst{1}}

\institute{
  \inst{1} Dipartimento di Fisica e Astronomia, Universit\`a di
Catania, and INFN sezione di Catania, \\ Via S. Sofia 64, I-95123
Catania, Italy, EU
}

   { High dimensional  chaos }
   { Synchronization }
  { Long-range interactions }
  { Phase transitions in statistical mechanics and thermodynamics}

\abstract{We study the chaotic behavior of the  synchronization phase transition 
in  the Kuramoto model. 
We discuss the  relationship with analogous features found  in the Hamiltonian 
Mean Field (HMF) model. Our numerical results support the connection between the two models, 
which can be considered as 
limiting cases (dissipative and conservative, respectively) of a more general dynamical system 
of damped-driven coupled pendula. We also show that, in the Kuramoto model, the shape of the phase 
transition and the largest Lyapunov exponent behavior are strongly dependent on the distribution 
of the natural frequencies.
}

\begin{document}

\maketitle

\section{Introduction}

Long-range interacting systems have been intensively studied 
in the last years and new methodologies have been developed
in the attempt to understanding their intriguing features.
One of the most promising directions is the combination of 
statistical mechanics tools and methods adopted in dynamical systems 
\cite{long-range}. 
In particular, phase transitions have been extensively  
explored in both conservative and dissipative long-range systems.
The Hamiltonian Mean Field (HMF) model \cite{hmf} and the Kuramoto model \cite{kuramoto,strogatz,acebron}
represent two paradigmatic toy models, the former conservative and  the latter dissipative,
for many real systems with long-range forces and have several applications.  
Both  models share the same order parameter and display a spontaneous phase transition 
from an homogeneous/incoherent phase to a magnetized/synchronized one. 
\\
In \cite{kura-hmf} 
we already observed that HMF and Kuramoto models can be considered as limiting cases (respectively 
conservative and overdamped) of a more general model of driven-damped coupled inertial oscillators. 
In this paper we present new numerical results which support a common
scenario for the two models.
More precisely, first 
we discuss the well known equilibrium features of the second order phase transition 
in the HMF model, then we study the stationary asymptotic behavior of the Kuramoto
model as a function of the coupling strength. 
On one hand, through new numerical simulations of large size systems,  
we confirm that, as also pointed out by other authors \cite{pazo} , the shape of the dynamical phase 
transition in the Kuramoto model changes from a continuous to an abrupt one, depending 
on the distribution of the natural frequencies of the oscillators.
On the other hand, and this is our main result, we clearly show  that, as  for the  HMF 
model, the largest Lyapunov exponent (LLE) of  the Kuramoto model
exhibits a peak just around the critical value, confirming the generality of
this microscopic signature for a phase transition.
Chaotic behavior in the Kuramoto model was discussed previously in ref.\cite{maistrenko}. However 
 those authors compute the entire spectrum of the Lyapunov exponents  only for small sizes and 
only for one kind of natural frequencies distribution, without discussing the strong dependence 
of the chaotic behavior on the shape of that distribution and its persistence in the thermodynamical limit.
Finally, by tuning the width of the natural frequencies distribution, we 
 show how the phase transition changes continously from 2nd-order-like behavior
towards  a 1st-order-like one 
and we draw a complete
synchronization phase diagram. As far as we know,  these results 
are reported for the first time and we think that they could provide  with  new insights
for the study of dynamical phase transitions in systems displaying collective synchronization.

\section{Phase transition and chaos in the HMF model}

The Hamiltonian Mean Field model describes the dynamics of $N$ classical spins or rotators,
characterized by the angles $\theta_{i}\in[-\pi,\pi[$ and the coniugate
momenta $p_{i}\in]-\infty,\infty[$, which can also be represented as particles moving on the unit circle. 
In its ferromagnetic version the Hamiltonian of the model is given by:

\begin{equation}
H=K+V=\sum_{i=1}^{N}\frac{p_{i}^{2}}{2m}+\frac{1}{2N}\sum_{i,j=1}^{N}
\left[1-\cos(\theta_{i}-\theta_{j})\right],
\label{hmf1}
\end{equation}

where $i=1,...,N$ and the mass $m$ is usually set to $1$.
The potential term of Eq.\ref{hmf1} reveals the mean field nature of the model,
since each rotator can interact with all the others. Such a nature becomes more evident
if we define as order parameter the magnetization 
$\mathbf{M}=Me^{i\phi}=\frac{1}{N}\sum_{j=1}^{N}e^{i\theta_{j}}$,
where $M$ and $\phi$ are the modulus and the global phase. 
Within this assumption the Hamilton equations of motion can be written 
\begin{equation}
\ddot{\theta_{i}}=\frac{1}{N}\sum_{j=1}^{N}\sin(\theta_{j}-\theta_{i})=
M\sin(\phi-\theta_{i}), \:\:\:\:\: i=1,\,...,N,
\label{eq:eq-moto-hmf}
\end{equation}
which correspond to the equations of single pendula in a mean field potential.
We note also \cite{kura-hmf} that Eq.\ref{eq:eq-moto-hmf} can be regarded as the conservative limit 
of the following mean field equation describing a system of driven and damped pendula (with unit mass): 
\begin{equation}
\ddot{\theta_i}+B\dot{\theta_i}+CM\sin(\theta_i - \phi) =\Gamma, \:\:\:\:\: i=1,\,...,N,
\label{eq-pendula}
\end{equation}
provided that the coupling $C=1$, the damping coefficient $B=0$ and the torque term $\Gamma=0$.     
\\
The equilibrium solution of the HMF model can be derived in both the canonical
and microcanonical ensembles \cite{hmf}. It gives the exact expression of 
the so-called {\it caloric curve}, i.e. the dependence of the energy density $U=H/N$ on the temperature $T$: 
$U=\frac{1}{2\beta}+\frac{1}{2}(1-M^{2})$,
being $\beta=1/T$, and predicts a second-order phase transition
from a low-energy condensed phase (with $M>0$) to an high-energy homogeneous one 
(with $M=0$) at the critical temperature $T_{C}=1/2$ (corresponding to the
critical energy density $U=4/3$).

\begin{figure}
\begin{center}
\includegraphics [scale=0.37] 
{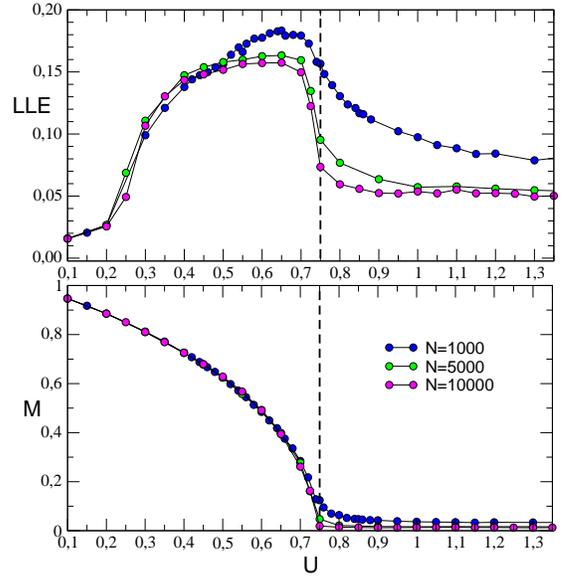}
\caption{ (Lower panel) The modulus of the order parameter $M$ of the HMF model is plotted as a function of the energy density $U$ for various system sizes at equilibrium: $N=1000, 5000,10000$; (Upper panel) Numerical calculation of the equilibrium largest Lyapunov exponent (LLE) as a function of $U$ for the same system sizes. In both the panels each dot represents an average over $10$ runs. See text. }
\label{fig1}
\end{center}
\end{figure}

The microcanonical simulations at equilibrium confirm these predictions and also 
allow to get some information about the microscopic dynamics of the system \cite{hmf,hmf-equilibrium} .
It is well known that in classical many-particle systems macroscopic collective behavior
can coexist with chaos at the microscopic level. This feature is particularly
evident near a phase transition, where chaotic dynamics can induce non trivial 
time dependence in macroscopic quantities. In these cases it is worthwhile to study
the Largest Lyapunov exponent (LLE), which gives a sufficient condition for chaotic 
instability by measuring the asymptotic rate of exponential growth of vectors in tangent 
space. For this purpose one has to consider the limit: 
$\lambda=\lim_{t\rightarrow\infty}\frac{1}{t}\ln\frac{d(t)}{d(0)}$,
where $d(t)=\sqrt{\sum_{i}\left((\delta\theta_{i})^{2}+(\delta p_{i})^{2}\right)}$
is the Euclidean distance calculated from the infitesimal displacements
at time $t$. Then, in order to obtain the time evolution of $d(t)$, one
must integrate along the reference orbit the linearized equations
of motion following for example the procedure of ref. \cite{benettin}. 
\\
In the upper panel of Fig.1 we plot the LLE as a function of the 
energy density for increasing system sizes $N$, while in the lower panel 
the correspondent magnetization curve, exibiting the typical shape of a continuous 
phase transition, is reported for comparison. An average over $10$ realizations at 
equilibrium has been considered for each point.
As expected, in both the limits of small and large energies, where the system becomes integrable, the LLE goes to zero. On the other hand, just before the critical energy,
LLE  exhibits a peak which persists and becomes broader increasing the size $N$ (see also \cite{hmf-equilibrium}).
In particular, it has been already shown (see Fig.16 in \cite{hmf}) that the LLE 
is positive and N-independent just below the transition, while it goes to zero above it (as $N^{-1/3}$)  and also  for very small energy densities.
Such a behavior at equilibrium is strikingly correlated to 
the kinetic energy fluctuations \cite{hmf,hmf-equilibrium} and it is also in agreement with a  theoretical
formula relating the LLE with other dynamical quantities \cite{firpo}, see \cite{pettini1,pettini2} for the general theory.
In the next section we will show that analogous features can be found also
in an apparently different context, as that one of the Kuramoto model.

\begin{figure}
\begin{center}
\includegraphics [scale=0.32] 
{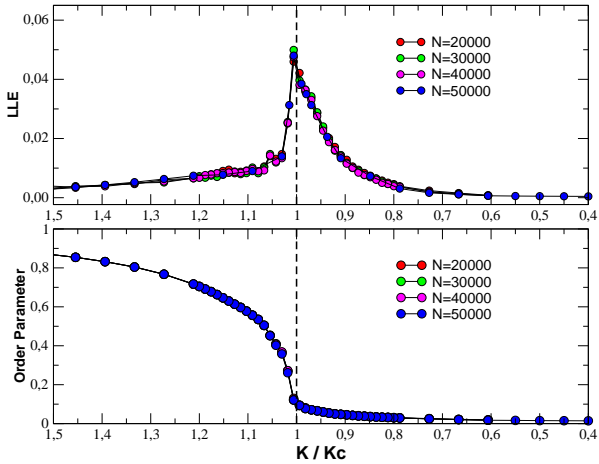}
\caption{(Lower panel) The asymptotic order parameter $r$ of the Kuramoto model is plotted as a function of the ratio $K/K_C$ for several system sizes ($N=20000,30000,40000$ and $50000$) and for a Gaussian distribution $g(\omega)$ of the natural frequencies. Decreasing values of $K/K_C$ are used in order to compare the data with the HMF ones. In this case $K_C=1.59617..$. A 2nd-order-like dynamical transition from the homogeneous phase to a synchronized one, similar to that one observed in the HMF model (Fig.1), is clearly visible. (Upper panel) The Largest Lyapunov Exponent (LLE) is plotted as function of $K/K_C$. 
A well defined peak around the phase transition indicates a microscopic chaotic activity which is maximal at the critical point}. In both  panels each dot represents an average over $10$ realizations. See text. 
\label{fig2}
\end{center}
\end{figure}

\section{Phase transition and chaos in the Kuramoto model}

The Kuramoto model \cite{kuramoto,strogatz,acebron} is considered one of the simplest models
exhibiting spontaneous collective synchronization. It describes
a large population of coupled limit-cycle oscillators, each one characterized by 
a phase $\theta_{i}$ and a natural frequency $\omega_{i}$, whose dynamics
is given by:
\begin{equation}
\dot{\theta}_{i}=\omega_{i}+\frac{K}{N}\sum_{j=1}^{N}\sin(\theta_{j}-\theta_{i}),
\label{eq:eq-moto-kura}
\end{equation}
where $K\geq0$ is the coupling strenght and $i=1,...N$. 
The natural frequencies are time-independent and
are randomly chosen from a symmetric, unimodal distribution $g(\omega)$.
We will consider here only uniform and Gaussian $g(\omega)$ distributions.
As in the case of HMF model, one can immagine the oscillators as particles moving
on the unit circle. For small values of $K$, the oscillators act
as if they were uncoupled and each oscillator tends to run independently
and incoherently with its own frequency. Instead, when $K$ exceeds a certain threshold
$K_{C}$, the coupling tends to synchronize each oscillator with all
the others and the system exhibits a spontaneous transition from the previous incoherent state 
to a synchronized one, where all the oscillators rotate at the same frequency $\Omega$ 
(a value which corresponds to the average frequency of the system, preserved  by the dynamics). 
As shown by Kuramoto itself \cite{kuramoto}, the critical value 
of the coupling depends only on the central value $g(\omega=0)$ of the distribution $g(\omega)$
in accordance with the expression $K_{C}=\frac{2}{\pi g(0)}$.

\begin{figure}
\begin{center}
\includegraphics [scale=0.32] 
{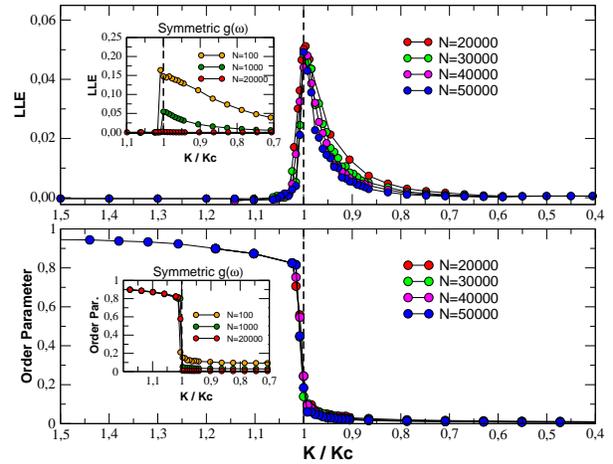}
\caption{(Lower panel) The asymptotic order parameter $r$ of the Kuramoto model is plotted as a function of the ratio $K/K_C$ for several system sizes ($N=20000,30000,40000$ and $50000$) and for a uniform distribution $g(\omega)$ of the natural frequencies. The critical coupling is in this case $K_C=8/\pi$. A 1st-order-like dynamical transition from the homogeneous phase to a synchronized one is clearly visible. This is more evident in the inset where a "symmetric" uniform distribution was used, see text. (Upper panel) The Largest Lyapunov Exponent (LLE) is plotted as function of $K/K_C$. A sharp peak around the phase transition indicates microscopic chaotic activity. Also in this case we show in the inset  the results for the "symmetric" frequency distribution, see text,  for which the transition is sharper. In all panels each dot represents an average over $10$ runs.
See text. } 
\label{fig2}
\end{center}
\end{figure}

The order parameter of the Kuramoto model is perfectly equivalent to the magnetizaton in the HMF model
and it is given by 
$\mathbf{r}=re^{i\phi}=\frac{1}{N}\sum_{j=1}^{N}e^{i\theta_{j}}$,
where $\phi$ is, again, the average global phase corresponding to the centroid of 
the phases of the oscillators and the modulus $0<r<1$ represents the degree of 
synchronization of the population. 
In terms of the variables $r$ and $\phi$, eq. (\ref{eq:eq-moto-kura})
can be rewritten as: 
\begin{equation}
\dot{\theta}_{i}=\omega_{i}+Kr\sin(\phi-\theta_{i}),\:\:\:\:\: i=1,\,...,N.
\label{eq:eq-moto-r}
\end{equation}
where, as happened also for the HMF model, the mean field character of the system becomes obvious.
For a given value of $K$, as the population becomes more coherent, $r$ grows and the effective
coupling $Kr$ increases. In this regime of partial synchronization, as predicted by the solutions of 
eq.(\ref{eq:eq-moto-r}), two kinds of oscillators coexist depending on the size of $\left|\omega_{i}\right|$ 
relative to $Kr$: (i)
oscillators with $\left|\omega_{i}\right|>Kr$,
called $drifting-oscillators$, that run incoherently around the unit circle;
(ii) oscillators with $\left|\omega_{i}\right|\leq Kr$, called $locked-oscillators$,  
that are trapped in a rotating cluster.     
The dynamic interplay between these two kinds of oscillators is probably at the root of the
microscopic chaotic behavior which, as we will show, characterizes the regime of
partial synchronization. On the other hand, when the effective coupling $Kr$ becomes strong enough, 
all the oscillators rotate in the same cluster at the frequency $\Omega$  
and any fingerprints of chaos disappears:  in fact, in this case,
the system behaves like a single giant oscillator and becomes thus integrable.
\\
If we consider again eq.(\ref{eq-pendula}) describing
a system of coupled driven/damped pendula, one can immediately verify that
eq.(\ref{eq:eq-moto-r}) represents its overdamped limit, i.e. the case $B>>1$ \cite{kura-hmf}. 
In this context, the natural frequencies $\omega_{i}$ play the role of the torque term, while $C=K$ and $M=r$.
This common origin of both HMF and Kuramoto models from eq.(\ref{eq-pendula}) seems to indicate 
the existence of a non trivial link between the two oscillators systems, despite the non-Hamiltonian character of 
the latter. Actually, their dynamics reveals many analogies. 
In refs. \cite{tanaka} the authors studied a generalized version of the Kuramoto model which share similarities with  the behavior  of the HMF model.
On the other hand,
in ref.\cite{kura-hmf} we  studied  analogies
in the quasi-stationary behavior, i.e. the appearance of metastable states near the phase transition.
In the following we will compare the stationary behavior of the Kuramoto model with the equilibrium
regime of the HMF model, with particular focus on the chaotic aspects.
\\
In the lower panels of Fig.2 and Fig.3 we show the asymptotic behavior
of the Kuramoto order parameter $r$ as a function of the control parameter $K$
for two different distributions of the natural frequencies $g(\omega)$,
respectively a Gaussian and a uniform  one, 
and for large systems of oscillators (from $N=20000$ to $N=50000$).
As predicted by Kuramoto analysis, in both  cases we observe
a phase transition at a critical value $K_{C}=\frac{2}{\pi g(0)}$.
The Gaussian distribution has mean $0$ and variance $1$, therefore 
from the normalization condition follows $g(0)=1/\sqrt(2\pi)$ and
$K_C=1.59617...$. 
The uniform distribution is selected in the range $\omega\in[-2,2]$ therefore the
normalization conditions gives $g(0)=1/4$ and $K_C=8/\pi$.
In order to compare the two cases, we plot
the order parameter as a function of $K/K_C$ and for several sizes 
of the system. Please notice that we show decreasing values of $K/K_C$ in order to better 
compare Kuramoto data with those of  the HMF model. An average over $10$  realizations
has been considered for each dot.
One immediately recognizes a different kind of transition: a continuous (2nd-order-like) one, for the Gaussian $g(\omega)$ (Fig.2) and an abrupt (1st-order-like) one, for the uniform $g(\omega)$ (Fig.3). Correspondingly, 
two different behaviors of the LLE (calculated as in the previous section following ref. \cite{benettin}) 
were also observed: they are plotted in the upper panels of Figs.2 and 3 and clearly show that 
 the LLE can be considered as a good dynamical indicator of the phase transitions. 
{In fact, in both  cases we observe a pronounced peak around the transition. But, while it slowly decreases for $K > K_C$ in the Gaussian $g(\omega)$ case, in the uniform one it  goes to zero abruptly
just after the critical point}. 
In both cases the chaotic regime, characterized by a positive finite LLE, seems to be related with 
the simultaneous presence of drifting and locking oscillators, i.e. with the existence of partially 
synchronized asymptotic stationary states, and seems not to depend on the size of the system
(as happened below the critical energy in the HMF model).
On the other hand, as expected, the LLE vanishes for small or high values of the coupling, 
being in those cases the system completely homogeneous or fully synchronized 
(i.e., in both the cases, integrable). 
\begin{figure}
\begin{center}
\includegraphics [scale=0.32] 
{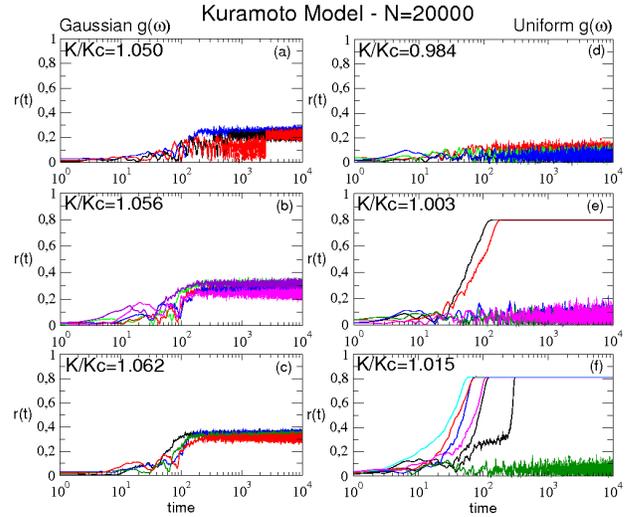}
\caption{Temporal evolution of the order parameter $r(t)$ near the phase transition for several runs and different       distributions $g(\omega)$: the Gaussian one, panels (a), (b) and (c) and the uniform one, panels (d), (e) and (f). 
Metastable states are visible in panels (a) and (f). See text.  } 
\label{fig3}
\end{center}
\end{figure}

These results confirm previous studies \cite{pazo} concerning  the
dependence of the transition order on the $g(\omega)$ distribution, and extend 
the investigation of  Maistrenko et al. \cite{maistrenko} without any contradiction. 
In particular, in the latter, the authors show that phase chaos in Kuramoto model arises 
as soon as $N=4$ or more oscillators interact. But, even if they compute the entire Lyapunov 
spectrum, indeed they take into account only relatively small system sizes (up to $N\le 200$) and do 
not distinguish between different kinds of phase transition. 
Furthermore, they mainly consider the so-called $symmetric$ 
Kuramoto model, where the natural frequencies $\omega_i$ are symmetrically allocated around the mean 
frequency $\Omega$: the latter is a very peculiar case, which gives rise to a very sharp
1st-order-like phase transition for the order parameter, with a corresponding LLE that is zero
for all the values of the coupling except for a very narrow zone around the phase transition,
which seems to vanish increasing the size of the system.  Numerical results 
for the symmetric $g(\omega)$ distribution are shown in the 
insets of Fig.3, where the sharp transition in the order parameter is clearly evident (lower inset),
together with the correspondent size-dependent LLE behavior (upper inset).
This distribution is however very peculiar and not very realistic, although 
easier to deal with from an analytical point of view.
{On the other hand, compared with those of Fig.1, the plots of Fig.2 seem to indicate 
that the Gaussian $g(\omega)$ choice for the natural frequencies
distribution yields a phase transition and a chaotic behavior qualitatively analogous
to that one found in the HMF model}.
%
\begin{figure}
\begin{center}
\includegraphics [scale=0.3] 
{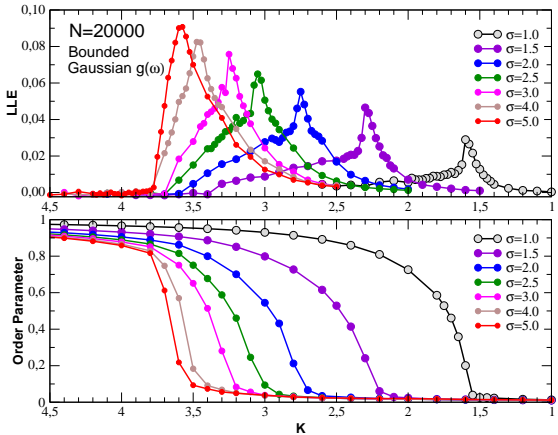}
\caption{(Lower panel) The asymptotic order parameter $r$ of the Kuramoto model is plotted as a function of the coupling $K$ for a system of $N=20000$ oscillators and for several bounded Gaussian distributions $g(\omega)$, with different standard deviations $\sigma$ and $\omega\in[-3,3]$.   
Increasing $\sigma$ (curves from right to left), the $g(\omega)$'s change continuously from a Gaussian to a uniform distribution and, in correspondence, the phase transition changes continuously from a 2nd-order-like to a 1st-order-like one, while the critical value of the coupling $K_C$ shift to the left. (Upper panel) The Largest Lyapunov Exponent is plotted as function of $K$.  Each dot represents an average over $10$ runs.
See text. } 
\label{fig3}
\end{center}
\end{figure}
Comparing Fig.2 and Fig.3, it clearly appears that, at variance with what happens in the uniform 
$g(\omega)$ case, where both homogeneous and synchronized states simultaneously appear in correspondence 
of the abrupt phase transition, the Gaussian $g(\omega)$ distribution drives the 
Kuramoto system along an HMF-like continuous transition without coexistence
of different phases.
In Fig.4 we present several plots which confirm this interesting feature. 
For a system of $N=20000$ oscillators, we draw the temporal evolution of 
the order parameter $r(t)$ for several single runs as a function of three
values $K/K_C$ near the phase transition, for both Gaussian (left column) and 
 uniform (right column) $g(\omega)$.
It clearly appears that in the latter case (panels (e) and (f)) stationary states with 
high and low asymptotic values of $r$ coexist, while in the former case (panels (a), (b) and (c)) 
only partially synchronized stationary states are visible.
Such a result reinforces the distinction between the 1st-order-like and 2nd-order-like
dynamical phase transitions,  occurring in the Kuramoto model in correspondence of different 
 $g(\omega)$ distributions, which seem to play a very crucial  role.
As already noticed in ref.\cite{kura-hmf}, in some cases (see for example panels (a) and (f)) 
also metastable states appear, for both $g(\omega)$ distributions, analogously to the appearence
of metastable quasistationary states (QSS) in the HMF model (when the system starts from out-of-equilibrium
initial conditions).
A more detailed study on these states and on their chaotic properties, also
in relationship with the violation of Central Limit Theorem and with the metastable regime of 
the HMF model \cite{hmf-clt}, is in preparation \cite{kura-hmf-2}.

\begin{figure}
\begin{center}
\includegraphics [scale=0.32] 
{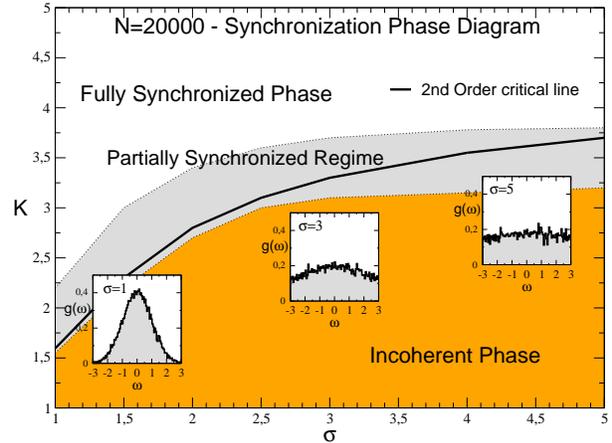}
\caption{For the Kuramoto model we show the phase diagram of $K$ versus $\sigma$ in the case with $N=20000$. A well defined critical line, changing asymptotically from 2nd-order-like towards 1st-order-like by increasing the value of $\sigma$, separates the fully synchronized phase from the incoherent one. The partially synchronized regime (characterized by a positive largest Lyapunov exponent and $0<r<0.8$) is also indicated (shaded area) between the two. In the insets, the bounded $g(\omega)$ distributions (with $\omega\in[-3,3]$) used in the simulations are plotted for three increasing values of $\sigma$. 
See text.  } 
\label{fig3}
\end{center}
\end{figure}
We conclude this section by showing that, although it is believed that for strictly unimodal $g(\omega)$ 
distributions the Kuramoto phase transition should be 2nd-order-like \cite{pazo}, 
the shape of the transition asymptotically tends to a 1st-order-like one
when  using bounded Gaussian distributions with an increasing standard deviation $\sigma$
and $\omega\in[-3,3]$.  We have adopted this range, instead of the $\omega\in[-2,2]$ previously 
used for the uniform distribution, in order to have a suitable Gaussian for small value of $\sigma$. 
In such a way one obtains, for $\sigma=1$, the Gaussian distribution used in Fig.2, which becomes larger and larger  by increasing the standard deviation. For $\sigma=5$, one obtains an almost uniform distribution,
like that used in Fig.3 (see for example the insets of Fig.6). 
By calculating the order parameter and the LLE as function of $K$ for several values of $\sigma$,
we obtain the curves shown in Fig.5 for $N=20000$. In this figure a continuous transformation from 
a 2nd-order HMF-like phase transition (lower panel, on the right) towards an abrupt 1st-order-like 
one (on the left) is clearly visible, 
{together with an analogous change in 
the peak of the largest Lyapunov exponent (upper panel)}, whose maximum value is also related with $g(\omega)$. 
In correspondence, the critical value $K_C$, initially depending on $\sigma$, moves from right to left 
towards a final value which does not depend on $\sigma$ any more 
and is very near to the theoretical value $K_C=12/\pi$ 
predicted for a true uniform $g(\omega)$ with $\omega\in[-3,3]$ (which in principle would 
strictly request $\sigma=\infty$).
At the same time, the region of partial sinchronization, which is mainly situated after the
phase transition for small values of $\sigma$ (see Fig.2), progressively shifts before the 
phase transition for increasing values of $\sigma$, approaching the 1st-order-like behavior 
shown in Fig.3.
This scenario is summarized in the plot of Fig.6, where the synchronization phase diagram of 
$K$ versus $\sigma$ for the Kuramoto model is shown. 
Please notice that this diagram  is schematic and not universal since it depends on the range 
of $g(\omega)$ and is likely affected also by unavoidable finite size effects.
We report in the three insets  examples of $g(\omega)$ distributions for $\sigma=1,3,5$. 
The 2nd-order-like critical line, drawn as a full line, separates the incoherent phase, with vanishing values for both $r$ and the LLE, from the fully synchronized one, characterized by a large value of the order parameter ($r>0.8$) and, again, by a vanishing LLE. 
Just around the critical line we found the partially synchronized regime, with positive LLE  and values $0<r<0.8$.
As  a final remark, we notice that in Ref.\cite{antoniazzi} a similar phase diagram was shown for  the HMF model.
In that case the authors considered the plane $U$ versus $M_0$, being the latter a parameter which specify 
the class of out-of-equilibrium initial conditions leading to metastable quasistationary states. Such a plane was
separated into two parts by a critical line, indicating both 2nd-order and 1st-order phase transitions
from an homogeneous QSS regime to a magnetized one. Despite the different context, we think that this analogy 
could be considered a further point of contact between the HMF and the Kuramoto scenarios.

\section{Conclusions}

We presented new numerical evidence of the presence of  chaotic behavior 
in the Kuramoto model for very large system sizes, discussing the 
analogies with the Hamiltonian Mean Field (HMF) model.
We studied the phase transition features and the LLE 
behavior for both  models. The latter can be also regarded as the dissipative
and the conservative version of a more general model of coupled driven/damped
pendula. 
Our simulations confirm that two different kinds of dynamical phase transitions 
occur in the Kuramoto model, depending on the distribution of the natural 
frequencies adopted as driving term. 
A uniform $g(\omega)$ gives rise to a sharp 1st-order-like
transition, where both homogeneous and synchronized stationary states coexist.
Instead, a Gaussian $g(\omega)$ yields a continuous 2nd-order-like
transition, very similar to the true thermodynamical phase transition observed 
in the HMF model.
On the other hand, the presence in the Kuramoto model of a peak observed in the 
LLE correspondingly to the critical region and regardless of the kind of distribution 
$g(\omega)$ reflects the fact that in this region the competition between locked
and drifting oscillators activates a microscopic chaotic dynamics which is a good
dynamical indicator of the phase transition.
Again, such a chaotic behavior shows many analogies with the one 
observed in the HMF model, which exhibits as well a peak just before the critical point, 
where there are large fluctuations in the main thermodynamical quantities 
characterizing the macroscopic phase transition.

\acknowledgments
We  thank  Marcello Iacono Manno for help in the preparation of the  
scripts  to run  our  codes on the TRIGRID platform. 
Useful discussions with {Antonio Politi,} Stefano Ruffo and Duccio Fanelli are acknowledged.

\end{document}